\documentclass[12pt,journal,onecolumn]{IEEEtran}
\usepackage{amsmath,amsfonts}
\usepackage{amssymb}
\usepackage{algorithmic}
\usepackage{algorithm}
\usepackage{array}
\usepackage[caption=false,font=normalsize,labelfont=sf,textfont=sf]{subfig}
\usepackage{textcomp}
\usepackage{stfloats}
\usepackage{url}
\usepackage{verbatim}
\usepackage{graphicx}
\usepackage{cite}
\usepackage{float} 
\usepackage{subcaption}
 \usepackage{multirow}
 \usepackage{caption}
\usepackage{threeparttable}
\usepackage{amsthm}
\usepackage{arydshln}
\usepackage{amssymb}
\usepackage{amsmath}
\usepackage{longtable}
\usepackage{array}
\usepackage{graphicx}
 \usepackage{multirow}
\usepackage{color}
\usepackage{enumerate}
\usepackage{booktabs}
\usepackage{longtable}
\usepackage[latin1]{inputenc}

\newtheorem{theorem}{\bf Theorem}
\newtheorem{lemma}{Lemma}

\newtheorem{remark}{Remark}

\newtheorem{definition}{Definition}
\newcommand{\Tr}{{\mathrm{Tr}}}
\newcommand{\gf}{{\mathbb{F}}}

\begin{document}

\title{Self-Orthogonal Minimal Codes from $n$-to-$1$ Mappings}

\author{Qin Du, Ziling Heng and Jun Zhang
\thanks{Q. Du and Z. Heng are with the School of Science, Chang'an University, Xi'an 710064, China,  and also with the State Key Laboratory of Integrated Services Networks, Xidian University, Xi'an 710071, China; J. Zhang is with the College of Mathematics and Statistics, Chongqing University, Chongqing 401331, China (e-mail: 15535565189@163.com, zilingheng@chd.edu.cn, jun\_zhang@cqu.edu.cn). \emph{(Corresponding author: Ziling Heng)}}
\thanks{Z. Heng's research was supported in part by the National Natural Science Foundation of China under Grant 12271059 and in part by the Natural Science Basic Research Program of Shaanxi (Program No. 2026JC-YXQN-030). J. Zhang's research was supported by the National Natural Science Foundation of China under the Grant No. 12441105.}}



\maketitle

\begin{abstract}
Self-orthogonal codes are central to the construction of quantum error-correcting codes, and minimal codes yield secret-sharing schemes whose access structures are read off from the supports of their codewords. A code that is both self-orthogonal and minimal therefore serves both purposes at once, so that a single family of codes can produce a quantum code and a secret-sharing scheme simultaneously. Linear codes arising from $n$-to-$1$ mappings over finite fields are a natural source of such codes and have attracted considerable attention recently, yet their parameters and weight distributions remain difficult to determine: for most high-degree functions the Walsh spectra cannot be computed directly, and it is precisely these weights on which the analysis of minimality and self-orthogonality rests. We address this difficulty through the algebraic theory of skew polynomial rings. Using the right greatest common divisor of skew polynomials, we determine the dimensions of the kernel spaces of the linearized polynomials associated with the $n$-to-$1$ mappings, which in turn yield the weight distributions of the corresponding codes. With this method we treat two newly constructed families of $n$-to-$1$ functions together with one known family, and obtain the exact weight distributions of the resulting codes; we further prove that all of these codes are minimal and self-orthogonal. Each family therefore simultaneously yields a quantum error-correcting code and a secret-sharing scheme, with explicitly determined parameters and access structures. This work offers a new perspective for determining the parameters of linear codes derived from $n$-to-$1$ mappings.
\end{abstract}


\section{Introduction}\label{sec1}

A linear code of length \(v\) and dimension \(\kappa\) over the finite field \(\mathbb{F}_{q}\) is defined as a \(\kappa\)-dimensional linear subspace of \(\mathbb{F}_{q}^{v}\). It is customarily referred to as a \([v,\kappa,d]\) code, where \(d\) denotes the minimum Hamming distance.
The weight distribution of a \([v,\kappa,d]\) linear code \(\mathcal{C}\) is the sequence \(\left(A_{i}\right)_{i=0}^{v}\), where \(A_{i}\) denotes the number of codewords of Hamming weight \(i\). It is a fundamental invariant that not only determines the error-detecting and error-correcting capabilities of the code but is also closely related to many of its combinatorial and algebraic properties.

With the rapid development of modern communications, data storage, distributed computing, and cryptographic systems, the demand for efficient and reliable coding schemes has become increasingly urgent. Linear codes over finite fields have been widely studied due to their linear structure and ease of implementation. Linear codes with special weight distributions and good orthogonality properties have always been a focus of research. Few-weight linear codes have found diverse applications in cryptography and combinatorial designs \cite{scheme, JC}. Consequently, the construction of new linear codes and the determination of their weight distributions are of both theoretical and practical interest.

Over the past decades, various methods for constructing linear codes have been proposed, yielding many families of codes with desirable parameters and interesting weight distributions \cite{scheme, Heng1, threecode, KC, KC2, Heng, threew}. In particular, researchers have constructed linear codes with optimal parameters over finite fields by employing various special cryptographic functions, such as (weakly regular) bent functions \cite{threew}, quadratic cryptographic functions \cite{TC2,ZLN}, weakly regular plateaued functions \cite{SM2,SM3}, (almost) perfect nonlinear (APN or PN) functions \cite{KC}, and \(n\)-to-1 functions \cite{q2,2y}, among others. While the core challenge in this approach lies in selecting suitable functions to construct the codes, the fundamental objective of such selection is to endow the resulting codes with desirable algebraic and combinatorial properties. Among these properties, minimality and self-orthogonality have emerged as two highly active research topics, owing to their wide applications in cryptography, communication systems, and data storage.

Minimal codes are characterized by the property that no codeword covers another, making them essential for constructing secret sharing schemes and secure multiparty computation protocols \cite{JC}. Since the introduction of the Ashikhmin-Barg sufficient condition \cite{A1}, the criteria for minimal codes have been extensively investigated; thereafter, the focus gradually shifted toward complete characterizations. In this direction, Heng, Ding and Zhou established a necessary and sufficient condition for linear codes over finite fields to be minimal via the Hamming weights of codewords, and constructed several families of binary and ternary minimal codes based on this condition \cite{Heng3}. From a different perspective, Tang et al. provided a complete characterization of minimal codes through the lens of projective geometry by employing cutting blocking sets \cite{TC3}. These contributions collectively advanced the theoretical framework of minimality criteria.

Meanwhile, self-orthogonal codes form a crucial subclass of linear codes. They are closely related to combinatorial designs, quantum error-correcting codes \cite{C}, linear complementary dual codes \cite{Heng4}, and secret sharing schemes \cite{Cruz}. Consequently, a great deal of research effort has been devoted to exploring their structural features and explicit constructions over the past decades. For instance, self-orthogonal codes  from combinatorial designs were investigated  \cite{DMGI}. More recently, researchers  constructed self-orthogonal codes by using special cryptographic functions, such as weakly regular bent functions and plateaued functions \cite{Heng4, WJK}.

Among known mathematical tools, $n$-to-$1$ mappings over finite fields are important for constructing good linear codes in coding theory. To the best of our knowledge, the study of constructing linear codes via $2$-to-$1$ functions dates back to \cite{Ding1,Ding2}. Following an important work in \cite{Qu1}, Li et al. further studied 2-to-1 mappings over finite fields of characteristic 2 and applied them to construct binary linear codes with new parameters via Construction I and Construction II \cite{KS}.
These two construction methods for linear codes are recalled as follows. Let $\mathbb{F}_{2^m}$ be the finite field with $2^m$ elements and $\mathbb{F}_{2^m}^*:=\mathbb{F}_{2^m}\setminus\{0\}$.
Let $f(x)$ be a function from $\mathbb{F}_{2^m}$ to $\mathbb{F}_{2^m}$ and $D(f)=\left\{f(x):x\in\mathbb{F}_{2^m}\right\}\setminus\{0\}=\left\{d_{1},d_{2},\cdots,d_{l}\right\}$, where $l=|D(f)|$. Let $\mathrm{Tr}_{2^m/2}$ be the trace function from $\mathbb{F}_{2^m}$ to $\mathbb{F}_{2}$.
Construction I is given by
\begin{eqnarray*}
\mathcal{C}_{f}=\left\{\mathbf{c}_{a,b}=\left(\mathrm{Tr}_{2^m/2}(ax+bf(x))\right)_{x\in \mathbb{F}_{2^m}^{*}}:a,b\in \mathbb{F}_{2^m}\right\}.
\end{eqnarray*}
Construction II is given as
\begin{eqnarray*}
\mathcal{C}_{D(f)}&=&\left\{\mathbf{c}_{b}=\left(\mathrm{Tr}_{2^m/2}\left(bd_{1}\right),\mathrm{Tr}_{2^m/2}\left(bd_{2}\right),\cdots,\mathrm{Tr}_{2^m/2}\left(bd_{l}\right)\right):b\in \mathbb{F}_{2^m}\right\}.
\end{eqnarray*}
In 2023, Mesnager et al. constructed several families of new binary minimal linear codes from 2-to-1 mappings
based on Construction III \cite{q2}, which is given by
\begin{eqnarray*}
\tilde{\mathcal{C}}_{D(f)}&=&\left\{\tilde{\mathbf{c}}_{a,b}=\left(\mathrm{Tr}_{2^m/2}(ad_{i}+bd_{j})\right)_{(d_{i},d_{j})\in D(f)\times D(f)}:a,b\in \mathbb{F}_{2^m}\right\}, 
\end{eqnarray*}
where $D(f)^2$ denotes the Cartesian product. Subsequently, in 2024, Ren\'e Rodr\'iguez-Aldama \cite{q1} extended this line of research to finite fields of arbitrary prime characteristic $p$, constructing two infinite families of $p$-ary linear codes via Construction II from $t$-to-$1$ functions.
More recently, Li et al.\cite{Lyq} constructed new $2$-to-$1$ functions of the form $f(x)=cx+\Tr_{2^{2m}/2^m}\left(h(x)\right)$ over $\mathbb{F}_{2^m}$, and obtained binary linear codes with new parameters via the Construction I, which enriched the research system in the direction.

Self-orthogonality and minimality serve distinct applications-quantum error-correcting codes and secret sharing schemes, respectively-yet linear codes possessing both properties are of considerable interest. A self-orthogonal code yields a quantum error-correcting code through the CSS construction \cite{C}, while a minimal code yields, through the construction method of Yuan and Ding \cite{JC}, a secret sharing scheme whose access structure is determined by the supports of its codewords; a code with both properties therefore supports both constructions at once. In settings that call for both quantum error correction and a specified access structure-CSS-based quantum secret sharing, for example, where a single self-orthogonal code provides the quantum code and the classical data from which the access structure is derived one code family then suffices in place of two. This motivates the construction of linear codes that are simultaneously self-orthogonal and minimal.

Consequently, the main purpose of this paper is to construct new families of such codes from $n$-to-$1$ mappings. Linear codes arising from $n$-to-$1$ mappings over finite fields are a natural source of such codes and have attracted considerable attention recently, yet their parameters and weight distributions remain difficult to determine: for most high-degree functions the Walsh spectra cannot be computed directly, and it is precisely these weights on which the analysis of minimality and self-orthogonality rests.

To address this issue, we adopt an approach based on the algebraic theory of skew polynomial rings. Specifically, by employing the right greatest common divisor method for polynomials over skew polynomial rings, we determine the dimensions of the kernel spaces of the linearized polynomials associated with the $n$-to-$1$ mappings. Building on this method, we then fully determine the exact weight distributions of the three families of linear codes based on Construction I, which arise from two newly constructed families of $n$-to-$1$ mappings and one known family. Furthermore, we prove that the resulting codes are both minimal linear codes and self-orthogonal linear codes with three nonzero weights. The three classes of self-orthogonal minimal codes constructed herein can be used in secret sharing schemes following the approach in \cite{JC}, as well as in quantum codes.

The rest of this paper is organized as follows. In Section II, we introduce the relevant preliminaries. In Section III, we construct two new families of $n$-to-$1$ mappings and then study the weight distributions, self-orthogonality, and minimality of the linear codes based on Construction I from these two new families of $n$-to-$1$ mappings as well as one known family. In Section IV, we summarize the main results of this paper and conclude the paper.

\section{Preliminaries}\label{sect2}
In this section, we present some preliminaries which will be used to give our main results.

We first define some notations as follows. 
\begin{enumerate}
\item[$\bullet$] $q$: a power of a prime $p$;
\item[$\bullet$] $\mathbb{F}_{q}$: the finite field with $q$ elements; 
\item[$\bullet$] $\mathbb{F}_{q^m}$: an $m$-degree finite extension of $\mathbb{F}_{q}$, where $m$ is a positive integer;
\item[$\bullet$] $\mathbb{F}_{q^m}^{*}$: the multiplicative group of $\mathbb{F}_{q^m}$;
\item[$\bullet$] $\mathbb{F}_{q^m}[x]$: the polynomial ring over the finite field $\mathbb{F}_{q^m}$;
\item[$\bullet$] $\times$: Cartesian product of sets;
\item[$\bullet$] $\#A$: the cardinality of a set $A$;
\item[$\bullet$] $\mathrm{Tr}_{q^{m}/q}(x)=\sum_{i=0}^{m-1}x^{q^i}$: the trace function from $\mathbb{F}_{q^m}$ onto $\mathbb{F}_{q}$;
\item[$\bullet$] $\ker(f(x))=\{x\in \gf_q:f(x)=0\}$: the kernel of  a homomorphism mapping $f$ over $\mathbb{F}_{q}$;
 \item[$\bullet$] $\dim_{\mathbb{F}_{2}}(K)$: the dimension of the $\mathbb{F}_{2}$-vector space $K$;
 \item[$\bullet$] $\operatorname{wt}(\textbf{c})$: the Hamming weight of a vector $\textbf{c}$;
 \item[$\bullet$] $d(\mathcal{C})$: the minimum distance of a code $\mathcal{C}$. 
\end{enumerate}

\subsection{$n$-to-$1$ Mappings}
The definition of $n$-to-$1$ mappings was given in \cite{N} as follows.
\begin{definition}\cite{N}\label{nto1}
Let \(A\) and \(B\) be finite sets. A function \(f:A\to B\) is called an \textbf{\(n\)-to-\(1\) mapping} provided that either of the following conditions holds:
\begin{enumerate}
    \item If \(n\) divides \(\#A\), then every element \(b\in B\) has either \(n\) or \(0\) preimages in \(A\).
    \item If \(n\) does not divide \(\#A\), then every element of \(B\), except for one unique element, has either \(n\) or \(0\) preimages in \(A\). This exceptional element of \(B\) has exactly \(\#A\bmod n\) preimages in \(A\).
\end{enumerate}
\end{definition}

Below, We recall a known $n$-to-$1$ mapping over finite fields in the following. 

\begin{lemma}\label{2to1}\cite{Lyq}
Let $k$ be an odd positive integer, $q=2^k$ and $c\in \mathbb{F}_{q^2}$ such that $c^{q-1}=\omega$, where $\omega^3=1$. Then $f(x)=cx+\Tr_{q^2/q}\left(x^6+x^{2q+4}\right)$ is $2$-to-$1$ over $\mathbb{F}_{q^2}$.
\end{lemma}

\subsection{Trace Functions}

The trace function from \(\mathbb{F}_{q^m}\) onto \(\mathbb{F}_{q}\) is defined by
\[
\Tr_{q^m/q}(\alpha) = \alpha + \alpha^q + \alpha^{q^2} + \cdots + \alpha^{q^{m-1}}.
\]
It is a surjective \(\mathbb{F}_{q}\)-linear mapping and a fundamental tool in coding theory, cryptography, and sequence design. Specifically, it has been widely utilized to construct cyclic codes, special classes of linear codes, cryptographic functions, and various sequences with low correlation.

\begin{lemma}\label{Tr}
The trace function from $\mathbb{F}_{q^m}$ to $\mathbb{F}_{q}$ satisfies the following properties:
\begin{enumerate}[(\arabic*)]
  \item $\Tr_{q^m/q}(\alpha+\beta)=\Tr_{q^m/q}(\alpha)+\Tr_{q^m/q}(\beta)$ for all $\alpha,\beta\in\mathbb{F}_{q^m}$;
  \item $\Tr_{q^m/q}(c\alpha)=c\Tr_{q^m/q}(\alpha)$ for any $c\in\mathbb{F}_{q}$ and $\alpha\in \mathbb{F}_{q^m}$;
  \item $\Tr_{q^m/q}(\alpha^{q})=\Tr_{q^m/q}(\alpha)$ for all $\alpha\in \mathbb{F}_{q^m}$;
  \item $\Tr_{q^m/q}(\alpha)=m \alpha$ for every element $\alpha$ belonging to $\mathbb{F}_{q}$.
\end{enumerate}
\end{lemma}

\subsection{Quadratic Functions and Walsh Transforms}
Several constructions of linear codes have been presented in the Introduction. In what follows, our discussion is based on Construction~I. This approach defines a linear code over \(\mathbb{F}_{2}\) using a function \(f:\mathbb{F}_{2^m}\rightarrow \mathbb{F}_{2^m}\) as follows:
\begin{eqnarray}\label{eq1}
 \mathcal{C}_{f}=  \left\{ \mathbf{c}_{a,b} =\left (\mathrm{Tr}_{2^m/2}(ax + bf(x))\right)_{x \in \mathbb{F}_{2^m}^{*}} \mid a, b \in \mathbb{F}_{2^m} \right\}, 
\end{eqnarray}
where \(f(0)=0\) and \(m\) is a positive integer. To investigate the parameters of this code, it suffices to determine the Walsh transform \(W_{f}(a,b)\), defined as
\[
W_{f}(a,b)=\sum_{x\in \mathbb{F}_{2^m}}(-1)^{\mathrm{Tr}_{2^m/2}\left(ax+bf(x)\right)}.
\]
Specially, we are interested in the value distribution of $W_{f}(a,b)$ if $f$ is a quadratic function, i.e. $f$ has algebraic degree 2.

\begin{lemma}\label{walsh}\cite{2y}
Let $\varphi(x):=\Tr_{2^m/2}\left(Q(x)\right)$, where $Q(x)$ denotes a quadratic function from $\mathbb{F}_{2^m}$ to $\mathbb{F}_{2^m}$ such that $Q(0)=0$. The associated symmetric bilinear mapping is defined by $B_{\varphi}(x,y)=\varphi(x+y)+\varphi(x)+\varphi(y)$. The kernel of $B_{\varphi}$, denoted by $V_{\varphi}$, constitutes a $\mathbb{F}_{2}$-subspace of $\mathbb{F}_{2^m}$ defined as:
\begin{eqnarray*}
V_{\varphi}=\left\{y\in \mathbb{F}_{2^m}: B_{\varphi}(x,y)=0\ \text{for}\ \text{all}\ x\in \mathbb{F}_{2^m}\right\}.
\end{eqnarray*}
The rank of $\varphi$ is defined as $Rank(\varphi)=m-\dim_{\mathbb{F}_{2}}\left(V_{\varphi}\right)$. Then $\varphi(y)=\Tr_{2^m/2}\left(Q(y)\right)$ is linear over $V_{\varphi}$ and
\begin{eqnarray*}
\Sigma_{x\in \mathbb{F}_{2^m}}(-1)^{\varphi(x)}=
\begin{cases}
\pm2^{\frac{m+d^{'}}{2}} \ if\ \varphi\ \text{vanishes}\ \text{on}\ V_{\varphi},\\
0\ \ \ \ \ \ \ \             \  \text{otherwise},
\end{cases}
\end{eqnarray*}
where $d^{'}=\dim\left(V_{\varphi}\right)$.
\end{lemma}

Let $f$ be a quadratic function from $\mathbb{F}_{2^m}$ to $\mathbb{F}_{2^m}$ such that $f(0)=0$.
Denote by $\varphi_{a,b}(x)=\Tr_{2^m/2}\left(ax+bf(x)\right)$ and $\varphi_{b}(x)=\Tr_{2^m/2}\left(bf(x)\right)$ for $a,b\in\mathbb{F}_{2^m}$. 
Note that $\varphi_{a,b}(x)$ and $\varphi_{b}(x)$ have the same associated symmetric bilinear mapping for any nonzero $b\in \mathbb{F}_{2^m}$.
Then Lemma \ref{walsh} yields the following result.

\begin{lemma}\label{walsh}\cite{2y} Let $f$ be a quadratic function from $\mathbb{F}_{2^m}$ to $\mathbb{F}_{2^m}$ such that $f(0)=0$.
Then we have
\begin{eqnarray*}
W_{f}(a,b)=\Sigma_{x\in \mathbb{F}_{2^m}}(-1)^{\varphi_{a,b}(x)}=
\begin{cases}
\pm2^{\frac{m+d_{b}}{2}} \ \text{if}\ \varphi_{a,b}\ \text{vanishes}\ \text{on}\ V_{\varphi_{b}},\\
0\ \ \ \ \ \ \ \ \ \ \           \ \text{otherwise},
\end{cases}
\end{eqnarray*}
where $d_{b}$ denotes the dimension of the kernel $V_{\varphi_{b}}$ of $\varphi_{b}$.
\end{lemma}

\subsection{The Skew Polynomial Ring With Automorphism $\sigma$}
\begin{definition}\cite{Wubaof}\label{skew}
Let $\sigma$ be the Frobenius automorphism of $\mathbb{F}_{q^k}$ over $\mathbb{F}_{q}$, and denote by $\mathbb{F}_{q^k}[x;\sigma]$ the set of all skew polynomials over $\mathbb{F}_{q^k}$. Addition of such polynomials is defined in the usual way, while multiplication is defined via the distributive law together with the rule
\begin{eqnarray}\label{eq2}
(ax^{i})(bx^{j})=a\sigma^{i}(b)x^{i+j}=ab^{q^{i}}x^{i+j},\ \ a,b\in \mathbb{F}_{q^k}.
\end{eqnarray}
With these operations, $\mathbb{F}_{q^k}[x;\sigma]$ forms a ring, called the skew polynomial ring over $\mathbb{F}_{q^k}$ with automorphism $\sigma$. 
\end{definition}

\begin{definition}\cite{Ore1}\label{gcrd}
When $F(x)=D_{1}(x)D_{2}(x)$, we shall call $D_{2}(x)$ a right-hand and $D_{1}(x)$ a left-hand divisor of $F(x)$. 
\end{definition}

Let $F(x)=a_{0}x^{m_{1}}+a_{1}x^{m_{1}-1}+\cdots+a_{m_{1}}$ and $G(x)=b_{0}x^{m_{2}}+a_{0}x^{m_{2}-1}+\cdots+b_{m_{2}}\in\mathbb{F}_{q^k}[x;\sigma]$ and $m_{1}\geq m_{2}$. The difference 
\begin{eqnarray}\label{eq3}
F_{1}(x)=F(x)-a_{0}\sigma^{m_{1}-m_{2}}(b_{0}^{-1})x^{m_{1}-m_{2}}G(x)
\end{eqnarray}
is of degree lower than $m_{1}$. It follows that one can always perform a right-hand division
\begin{eqnarray*}
F(x)=Q(x)G(x)+R(x),
\end{eqnarray*}
where $Q(x)$ is of degree $m_{1}-m_{2}$ and the degree of $R(x)$ does not exceed $m_{2}-1$. Applying the right Euclidean algorithm with the above right division with remainder 
yields the right greatest common divisor of \(F(x)\) and \(G(x)\), denoted by $\mathrm{gcrd}(F(x),G(x))$. 
Specifically, if
\begin{eqnarray*}
F(x)&=&Q_{0}(x)G(x)+F_{1}(x),\\
G(x)&=&Q_{1}(x)F_{1}(x)+F_{2}(x),\\
&\vdots&\\
F_{t-2}(x)&=&Q_{t-1}(x)F_{t-1}(x)+F_{t}(x),\\
F_{t-1}(x)&=&Q_{t}(x)F_{t}(x),
\end{eqnarray*}
then $\mathrm{gcrd}(F(x),G(x))=F_{t}(x)$.

\subsection{Linearized Polynomials}

Both in theory and in applications, the linearized polynomials introduced below are of great importance. A useful feature of these polynomials is the structure of their root set, which facilitates the determination of the roots. As usual, let \(q\) denote a prime power.

\begin{definition}\cite{lidl}\label{linear1}
A polynomial of the form 
\begin{equation*}
L(x)=\sum_{i=0}^{r}\alpha_{i}x^{q^{i}},
\end{equation*}
with coefficients in an extension field \(\mathbb{F}_{q^m}\) of \(\mathbb{F}_{q}\), is called a $q$-linearized polynomial, or a \(q\)-polynomial, over \(\mathbb{F}_{q^m}\).
\end{definition}

\begin{definition}\cite{lidl}\label{linear2}
The polynomials
\begin{equation}\label{eq4}
L(x)=\sum_{i=0}^{r}\alpha_{i}x^{q^{i}} \quad \text{and} \quad l(x)=\sum_{i=0}^{r}\alpha_{i}x^{i} 
\end{equation}
over \(\mathbb{F}_{q^m}\) are called \(q\)-associates of each other. More specifically, \(l(x)\) is the conventional \(q\)-associate of \(L(x)\), and \(L(x)\) is the linearized \(q\)-associate of \(l(x)\).
\end{definition}

\begin{lemma}\cite[pp. 112--114]{lidl}\label{root}
Let \(L(x)\) be a linearized polynomial over \(\mathbb{F}_{q^m}\). Denote its kernel by 
\begin{equation*}
K=\ker L=\left\{x\in \mathbb{F}_{q^m} \mid L(x)=0\right\}.
\end{equation*}
The kernel \(K\) forms an \(\mathbb{F}_{q}\)-linear subspace. Suppose \(\dim_{\mathbb{F}_{q}}K=t\). Then the cardinality of \(K\) satisfies \(|K|=q^t\). Furthermore, the following properties hold.
\begin{enumerate}
  \item If \(\alpha\) belongs to the image set \(\operatorname{Im}(L)\) of \(L\), then the equation \(L(x)=\alpha\) has exactly \(q^t\) solutions.
  \item If \(\alpha\) is not contained in \(\operatorname{Im}(L)\), then the equation has no solution.
\end{enumerate}
\end{lemma}

The following lemma provides a useful tool for determining the number of roots of a linearized polynomial.

\begin{lemma}\cite{Wubaof}\label{wubaof}
For any linearized polynomial \(L(x)\in\mathbb{F}_{q^m}[x]\), denote its associated polynomial by \(l(x)\). Then the number of roots of \(L(x)\) in \(\mathbb{F}_{q^m}\) equals \(q^{\ell}\) if and only if \(\deg(\mathrm{gcrd}(l(x),x^m-1))=\ell\).
\end{lemma}

\subsection{Self-Orthogonal Codes and Minimal Linear Codes}
For a prime power \(q\) and a positive integer \(n\), the (Hamming) distance between two vectors \(\boldsymbol{c} = (c_1, \dots, c_v)\) and \(\boldsymbol{c}' = (c_1', \dots, c_v') \in \mathbb{F}_q^v\) is the number of coordinates in which they differ. A \(q\)-ary \([v, \kappa, d]\) code \(\mathcal{C}\) is a \(\kappa\)-dimensional subspace of \(\mathbb{F}_{q}^v\) with minimum distance \(d\). Vectors in \(\mathcal{C}\) are called codewords. The dual code of \(\mathcal{C}\), denoted by \(\mathcal{C}^{\perp}\), is defined as
\[
\mathcal{C}^{\perp} := \left\{ x \in \mathbb{F}_{q}^v \mid x \cdot y = 0 \text{ for all } y \in \mathcal{C} \right\},
\]
which is an \([v, v-\kappa]\) code, where \(\cdot\) denotes the standard dot product.
If $\mathcal{C}\subseteq\mathcal{C}^{\bot}$, then $\mathcal{C}$ is called a self-orthogonal code, that is, $\mathbf{a}\cdot\mathbf{b}=0$ for any codewords $\mathbf{a},\ \mathbf{b}\in \mathcal{C}$. In particular, if $\mathcal{C}=\mathcal{C}^{\bot}$, then $\mathcal{C}$ is said to be self-dual. 
The following gives a sufficient condition for a linear code to be self-orthogonal.

\begin{lemma}\cite{WV}\label{self}
Let $\mathcal{C}$ be a binary linear code. If every codeword in $\mathcal{C}$ has Hamming weight divisible by $4$, then $\mathcal{C}$ is self-orthogonal, i.e., $\mathcal{C}\subseteq\mathcal{C}^{\perp}$.
\end{lemma}

The \emph{support} of a vector \(\mathbf{u}=(u_1,u_2,\ldots,u_v) \in \mathbb{F}_{q}^v\), denoted by \(\mathrm{Supp}(\mathbf{u})\), is defined as the set of indices of its nonzero coordinates, i.e., \(\mathrm{Supp}(\mathbf{u}) = \{1\leq i \leq v\mid u_i \neq 0\}\). A vector \(\mathbf{u}\) is said to \emph{cover} another vector \(\mathbf{v}\) if \(\mathrm{Supp}(\mathbf{v}) \subseteq \mathrm{Supp}(\mathbf{u})\), denoted by \(\mathbf{v} \preceq \mathbf{u}\). 
A codeword \(\mathbf{c}\) in a linear code \(\mathcal{C}\) is called \emph{minimal} if the only codewords it covers are its scalar multiples \(a\mathbf{c}\) with \(a \in \mathbb{F}_{q}\). A linear code \(\mathcal{C}\) is called \emph{minimal} if every codeword in \(\mathcal{C}\) is minimal. The following theorem provides a sufficient condition for a linear code to be minimal.
\begin{lemma}\label{min}\cite{Heng1}
 Let \(\mathcal{C}\) be a linear code over \(\mathbb{F}_{q}\). Let \(w_{\mathrm{min}}\) and \(w_{\mathrm{max}}\) denote the minimum and maximum Hamming weights of nonzero codewords in \(\mathcal{C}\), respectively. If
\begin{eqnarray}\label{eq5}
\frac{w_{\mathrm{min}}}{w_{\mathrm{max}}}>\frac{q-1}{q},
\end{eqnarray}
then \(\mathcal{C}\) is minimal.
\end{lemma}

\subsection{The Pless Power Moments}
For an $[v,\kappa,d]$ linear code $\mathcal{C}$ over $\mathbb{F}_{q}$, the weight of $\mathbf{c}_{i}=\left(c_{1},c_{2},\cdots,c_{v}\right)$ is defined to be the number of nonzero $c_{i}$. Denote by $A_{i}$ the number of codewords with weight $i$ in $\mathcal{C}$, and by $B_{i}$ the number of codewords with weight $i$ in $\mathcal{C}^{\bot}$. The first four Pless power moments are presented for $q=2$ as follows \cite{WV}:
\begin{eqnarray*}
\sum_{j=0}^{v}A_{j}&=&2^{\kappa},\\
\sum_{j=0}^{v}jA_{j}&=&2^{\kappa-1}\left(v-B_{1}\right),\\
\sum_{j=0}^{v}j^{2}A_{j}&=&2^{\kappa-2}\left(v(v+1)-2(v+1)B_{1}+2B_{2}\right),\\
\sum_{j=0}^{v}j^{3}A_{j}&=&2^{\kappa-3}\left[v\left(v^{2}+3v-4\right)-\left(3v^{2}+3v-6\right)B_{1} +(3v-6)B_{2}-6B_{3}\right].
\end{eqnarray*}
The Pless power moments provide a powerful tool for determining the minimum weight of the dual of a linear code.

\section{New families of linear codes from $n$-to-$1$ mappings}

In this section, we employ three classes of \(n\)-to-1 mappings, two of which are newly constructed and one of which is known, to construct three families of self-orthogonal minimal codes with three weights.

Let $f$ be a mapping from $\mathbb{F}_{2^m}$ to itself with $f(0)=0$. Let \(\mathcal{C}_{f}\) be the linear code defined in Equation (\ref{eq1}). 
Clearly, \(\mathcal{C}_{f}\) has dimension at most \(2m\). To determine the dimension of \(\mathcal{C}_{f}\), it suffices to compute the size of the set of pairs \((a,b)\) for which \(\mathrm{Tr}_{2^m/2}(ax+bf(x))\) vanishes for all \(x\in\mathbb{F}_{2^m}\). Equivalently, since \(f(0)=0\), we have \(\dim(\mathcal{C}_{f})=2m-\dim_{\mathbb{F}_{2}}(K_{1})\), where
\begin{eqnarray}\label{eq6}
K_{1}=\left\{(a,b)\in \mathbb{F}_{2^m}^{2}:W_f(a,b)=2^{m}\right\}.
\end{eqnarray}
The Hamming weight of a codeword $\mathbf{c}_{a,b}\in \mathcal{C}_{f}$ is given by  
\begin{eqnarray}\label{eqn-Walsh}
\nonumber \operatorname{wt}(\mathbf{c}_{a,b})&=&\#\left\{x\in \mathbb{F}_{2^m}^{*}:\mathrm{Tr}_{2^m/2}(ax+bf(x))=1\right\}\\
\nonumber&=&2^m-\#\left\{x\in \mathbb{F}_{2^m}:\mathrm{Tr}_{2^m/2}(ax+bf(x))=0\right\}\\
\nonumber&=&2^m-\frac{1}{2}\sum_{x\in \mathbb{F}_{2^m}}\sum_{y\in  \mathbb{F}_{2}}(-1)^{y\mathrm{Tr}_{2^m/2}(ax+bf(x))}\\
\nonumber&=&2^{m-1}-\frac{1}{2}\sum_{x\in \mathbb{F}_{2^m}}(-1)^{\mathrm{Tr}_{2^m/2}(ax+bf(x))}\\
&=&2^{m-1}-\frac{1}{2}W_{f}(a,b),
\end{eqnarray}
where the third equality holds due to the well-known orthogonality relation of additive characters \cite{lidl}.

In what follows, we study the minimum distance of the dual of \(\mathcal{C}_{f}\), showing that \(\mathcal{C}_{f}\) is always a projective code for $m>1$.
\begin{lemma}\label{dualdeg3}
Let $f(x)$ be a function from $\mathbb{F}_{2^m}$ to itself with $f(0)=0$ for $m>1$. 
Then the minimum distance of $\mathcal{C}_{f}^\perp$ satisfies $d\left(\mathcal{C}_{f}^{\perp}\right)\geq3$.
\end{lemma}

\begin{proof}
A vector $\boldsymbol{\omega}=(\omega_{x})_{x\in\mathbb{F}_{2^m}^{*}}$ belongs to
$\mathcal{C}_{f}^{\perp}$ if and only if
\begin{equation*}
\sum_{x\in\mathbb{F}_{2^m}^{*}}\omega_{x}\operatorname{Tr}_{2^m/2}(ax+bf(x))=0,
\quad \forall\, a,b\in\mathbb{F}_{2^m},
\end{equation*}
which is equivalent to
\begin{eqnarray} \label{eq-tr}
\mathrm{Tr}_{2^m/2}\left(a\sum_{x\in\mathbb{F}_{2^m}^*}\omega_{x}x+b\sum_{x\in\mathbb{F}_{2^m}^*}\omega_{x}f(x)\right)=0
\end{eqnarray}
for all $a,b \in \mathbb{F}_{2^m}$.
Note that $\omega_x\in \{0,1\}$. 
Suppose that $\mathcal{C}_{f}^{\perp}$ contains a codeword
of weight one. Then there exists $x_{0}\in\mathbb{F}_{2^m}^{*}$ such that
$\omega_{x_{0}}=1$ and all other entries vanish. Equation~(\ref{eq-tr}) then gives
\[
\operatorname{Tr}_{2^m/2}\bigl(ax_{0}+bf(x_{0})\bigr)=0
\text{ for all } a,b\in\mathbb{F}_{2^m}.
\]
Taking $b=0$, we obtain $\operatorname{Tr}_{2^m/2}(ax_{0})=0$ for all
$a\in\mathbb{F}_{2^m}$. Equivalently, $x_{0}\mathbb{F}_{2^m}\subseteq
\ker(\operatorname{Tr}_{2^m/2})$, which is impossible since
$\dim_{\mathbb{F}_{2}}\ker(\operatorname{Tr}_{2^m/2})=m-1< m$.
Now suppose that $\mathcal{C}_{f}^{\perp}$ contains a codeword of weight two,
supported on two distinct elements $x_{1},x_{2}\in\mathbb{F}_{2^m}^{*}$.
Equation~(\ref{eq-tr}) then yields
\[
\operatorname{Tr}_{2^m/2}\!\bigl(a(x_{1}+x_{2})+b(f(x_{1})+f(x_{2}))\bigr)=0,\
 \forall\, a,b\in\mathbb{F}_{2^m}.
\]
Taking $b=0$, we get $\operatorname{Tr}_{2^m/2}(a(x_{1}+x_{2}))=0$ for all
$a\in\mathbb{F}_{2^m}$. Since $x_{1}\neq x_{2}$, we have $x_{1}+x_{2}\neq 0$.
Hence $(x_{1}+x_{2})\mathbb{F}_{2^m}\subseteq\ker(\operatorname{Tr}_{2^m/2})$,
which is impossible because
$\dim_{\mathbb{F}_{2}}\ker(\operatorname{Tr}_{2^m/2})=m-1<m$.
Consequently, $\mathcal{C}_{f}^{\perp}$ contains no nonzero codeword of
weight less than three, and hence $d(\mathcal{C}_{f}^{\perp})\geq 3$.
\end{proof}

\subsection{The case that $f(x)=\Tr_{q^l/q}\left(x^{2q+2}\right)+cx$}
In this subsection, we study the linear code $\mathcal{C}_{f}$ for
$f(x)=\operatorname{Tr}_{q^{l}/q}\bigl(x^{2q+2}\bigr)+cx$, where $c\in \mathbb{F}_{q}^{*}$ and $q=2^{k}$
with $k$ an even positive integer and $l$ an odd positive integer. 
Firstly, we prove that $f(x)$ is a $4$-to-$1$ mapping or a permutation polynomial. 

\begin{theorem}\label{4to1}
Let $q=2^{k}$ with $k$ an even positive integer, and let $l$ be an odd
positive integer. Define
$f(x)=\operatorname{Tr}_{q^{l}/q}\bigl(x^{2q+2}\bigr)+cx$ for
$c\in\mathbb{F}_{q}^{*}$. Then the following
statements hold.
\begin{enumerate}
  \item If $c^{\frac{q-1}{3}}=1$, then $f(x)$ is a $4$-to-$1$ mapping over $\mathbb{F}_{q^l}$.
  \item If $c^{\frac{q-1}{3}}\neq1$, then $f(x)$ is a permutation polynomial of $\mathbb{F}_{q^l}$.
\end{enumerate}
\end{theorem}

\begin{proof}
We analyze the solutions of the equation $f(x)=e$, i.e.,
\begin{eqnarray}\label{eq7}
\Tr_{q^l/q}\left(x^{2q+2}\right)+cx=e 
\end{eqnarray}
for any element $e\in \mathbb{F}_{q^l}$.
Let $u:=cx+e$. Then Equation (\ref{eq7}) implies $u=\Tr_{q^l/q}\left(x^{2q+2}\right)\in \mathbb{F}_{q}$ and $x=\frac{u+e}{c}$ for $c\neq0$. Due to $u\in\mathbb{F}_{q}$ and $c\in\mathbb{F}_{q}^{*}$, then
\begin{eqnarray*}
x^{2q+2}=\left(\frac{u+e}{c}\right)^{2q+2}=\frac{u^4+\left(e^2+e^{2q}\right)u^2+e^{2q+2}}{c^4}.
\end{eqnarray*}
Substituting it into Equation \eqref{eq7} and applying Lemma \ref{Tr}, we obtain
\begin{eqnarray*}
u=\frac{u^4+\left(e^2+e^{2q}\right)u^2+e^{2q+2}}{c^4}=\frac{u^4}{c^4}+\Tr_{q^l/q}\left(e^2+e^{2q}\right)\frac{u^2}{c^4}+\frac{\Tr_{q^l/q}\left(e^{2q+2}\right)}{c^4}.
\end{eqnarray*}
Rearranging the above equation yields a quartic equation in $u$:
\begin{eqnarray*}
u^4+c^4u+\Tr_{q^l/q}\left(e^{2q+2}\right)=0.
\end{eqnarray*}
Since $u^{4}+c^{4}u$ is an $\mathbb{F}_{2}$-linearized polynomial,
Lemma~\ref{root} implies that the equation
$u^{4}+c^{4}u+\operatorname{Tr}_{q^{l}/q}\bigl(e^{2q+2}\bigr)=0$ admits only
two possibilities for its number of solutions: either it has no solution, or
its number of solutions equals that of the homogeneous equation
$u^{4}+c^{4}u=0$. Hence, in what follows, it suffices to study the solutions
of $u^{4}+c^{4}u=0$ in $\mathbb{F}_{q}$. 
Obviously, $u=0$ is a trivial solution of $u^4+c^4u=0$. 
For $u\in \mathbb{F}_{q}^*$, we have $u^{3}+c^{4}=0$, which has solutions if and only if $c^4$ is a cubic element in $\mathbb{F}_{q}^*$.
Since $k$ is even, we have $3\mid 2^k-1$, which implies the existence of a cubic subgroup in $\mathbb{F}_{q}^{*}$. Consequently, if $c^4$ is a cubic residue in $\mathbb{F}_{q}^*$, then the equation $u^3+c^4=0$ has three distinct nonzero solutions in $\mathbb{F}_{q}^*$; if $c^4$ is not a cubic residue in $\mathbb{F}_{q}^*$, then the equation $u^3+c^4=0$ admits no nonzero solutions. Let $N$ denote the number of distinct roots of the equation $u^4+c^4u+\Tr_{q^l/q}\left(e^{2q+2}\right)=0$ in $\mathbb{F}_{q}$.  Based on the above discussions, we derive
\begin{eqnarray*}
N=
\begin{cases}
  1 \ \ \ \ \text{if}\ c^{\frac{4(q-1)}{3}}\neq1,\\
  4 \ \ \ \ \text{if}\ c^{\frac{4(q-1)}{3}}=1.
\end{cases}
\end{eqnarray*}
Note that $c^{\frac{4(q-1)}{3}}=1$ if and only if 
Since the mapping $u=cx+e$ is an affine transformation, the number of solutions of $f(x)$ in $\mathbb{F}_{q^l}$ remains the same as above.
\end{proof}

\begin{remark}
In Theorem~\ref{4to1}, we assume that \(l\) is odd and $k$ is even.
If \(l\) is odd, \(c^{\frac{q-1}{3}}\neq1\), \(k\) is even, and \(k\not \equiv 0 \pmod{3}\), it has been proved in \cite[Theorem 3.4]{qu} that \(f(x)\) is a permutation of \(\mathbb{F}_{q^{l}}\).
However, 2) of Theorem~\ref{4to1} does not have the restriction \(k\not \equiv 0 \pmod{3}\).
In addition, we also studied the case where \(l\) is odd, \(k\) is even, and \(c^{\frac{q-1}{3}}=1\).
If \(l\) is even and \(c\in \mathbb{F}_{q^{2}}^*\), then \(f(x)\) is a permutation of \(\mathbb{F}_{q^{l}}\), which was proved in \cite[Theorem 3.1]{qu}.
\end{remark}

Now we study the parameters, weight distribution, minimality, and self-orthogonality of \(\mathcal{C}_{f}\), which is constructed from the \(n\)-to-1 mapping \(f(x)\). 
To this end, we require the following lemma.

\begin{lemma}\label{dim}
Let $m=lk$ and $q=2^k$, where $l$ is an odd positive integer and $k$ is a positive integer.
 Then we have
\begin{eqnarray*}
\dim_{\mathbb{F}_{2}}(\ker( L_{b}))=k,
\end{eqnarray*}
where $L_{b}(y):=\mathrm{Tr}_{2^m/2^k}(b)y^2+\mathrm{Tr}_{2^m/2^k}(b)y^{2q^2}$ with $\mathrm{Tr}_{2^m/2^k}(b)\neq 0$ is a $2$-linearized polynomial for $y\in \mathbb{F}_{2^m}$.
\end{lemma}

\begin{proof}
 The conventional \(2\)-associate of $L_{b}(y)$ is given by
$l_{a,b}(y)=\mathrm{Tr}_{2^m/2^k}(b)y+\mathrm{Tr}_{2^m/2^k}(b)y^{2k+1}$. 
According to Lemma \ref{wubaof}, we need to determine
\begin{eqnarray*}
\text{gcrd}\left(\mathrm{Tr}_{2^m/2^k}(b)y^{2k+1}+\mathrm{Tr}_{2^m/2^k}(b)y,y^{kl}-1\right).
\end{eqnarray*}
Let $B:=\mathrm{Tr}_{2^m/2^k}(b)$ and $l=2t+1$. Note that $y^{lk}-1=y^{lk}+1$.Then we have a right-hand division:
\begin{align*}
y^{kl}+1 &= \left(B^{-1}\right)^{2^{k(l-2)}-1}y^{k(l-2)-1}\left(By^{2k+1}+By\right)+ y^{k(l-2)}+1.
\end{align*}
Since the degree of \(y^{k(l-2)}+1\) is not less than \(2k+1\), we need to continue applying the right-hand division algorithm until the degree of the remainder does not exceed \(2k\).
We have
\begin{eqnarray*}
y^{k(l-2)}+1&=&\left(B^{-1}\right)^{2^{k(l-4)}-1}y^{k(l-4)-1}\left(By^{2k+1}+By\right)+y^{k(l-4)}+1,\\
y^{k(l-4)}+1&=&\left(B^{-1}\right)^{2^{k(l-6)}-1}y^{k(l-6)-1}\left(By^{2k+1}+By\right)+y^{k(l-6)}+1,\\
&\vdots&\\
y^{5k}+1&=&\left(B^{-1}\right)^{2^{3k}-1}y^{3k-1}\left(By^{2k+1}+By\right)+y^{3k}+1,\\
y^{3k}+1&=&\left(B^{-1}\right)^{2^{k}-1}y^{k-1}\left(By^{2k+1}+By\right)+y^{k}+1.
\end{eqnarray*}
We then obtain the final expression as
\begin{eqnarray*}
y^{kl}+1&=&\sum_{i=1}^{t}\left(B^{-1}\right)^{2^{k(l-2i)-1}}y^{k(l-2i)-1}\left(By^{2k+1}+By\right)+y^{k}+1.
\end{eqnarray*}
Similarly, we have
\begin{eqnarray*}
By^{2k+1}+By&=&By^{k+1}\left(y^k+1\right)+By^{k+1}+By,\\
By^{k+1}+By&=&By\left(y^k+1\right).
\end{eqnarray*}
Then we arrive at the conclusion that \[\operatorname{gcrd}\left(\operatorname{Tr}_{2^m/2^k}(b)y^{2k+1}+\operatorname{Tr}_{2^m/2^k}(b)y,\; y^{kl}-1\right)=y^k+1.\] Finally, the desired conclusion follows from Lemma~\ref{wubaof}.
\end{proof}

\begin{theorem}\label{4to1code}
Let $f(x)=\operatorname{Tr}_{q^l/q}\left(x^{2q+2}\right)+cx\in \mathbb{F}_{q^{l}}[x]$, where $q=2^k$, $l$ is odd, $k$ is a positive integer, and $c\in\mathbb{F}_{q}^{*}$.
Let $m=kl$. Then $\mathcal{C}_{f}$ is a binary linear code with parameters
\[
\left[2^{lk}-1,\; k(l+1),\; 2^{lk-1}-2^{\frac{lk+k}{2}-1}\right]
\]
and weight distribution in Table \ref{table1}. If $k(l-1)\geq4$, then $\mathcal{C}_{f}$ is minimal. If $k(l+1)\geq 6$, then $\mathcal{C}_{f}$ is self-orthogonal.
\end{theorem}

\begin{table}[h]
\centering{\caption{Weight distribution of $\mathcal{C}_{f}$ in Theorem \ref{4to1code}}\label{table1}}%
\begin{tabular}{@{}cc@{}}
\hline
Weight & Frequency  \\
\hline
$0$ & $1$ \\
  $2^{kl-1}-2^{\frac{kl+k}{2}-1}$ & $2^{\frac{kl+k}{2}-1}-2^{\frac{kl-k}{2}-1}+2^{kl-1}-2^{kl-k-1}$ \\
  $2^{kl-1}$ & $2^{kl-k}+2^{kl+k}-2^{kl}-1$ \\
  $2^{kl-1}+2^{\frac{kl+k}{2}-1}$ & $2^{\frac{kl-k}{2}-1}-2^{\frac{kl+k}{2}-1}+2^{kl-1}-2^{kl-k-1}$ \\
\hline
\end{tabular}
\end{table}

\begin{proof}
 By Equation (\ref{eqn-Walsh}), we first need to compute the Walsh transforms $W_{f}(a,b)$ for any $a,b\in \mathbb{F}_{2^m}$. We have
\begin{eqnarray*}
W_{f}(a,b)=\sum_{x\in \mathbb{F}_{2^{m}}}(-1)^{\mathrm{Tr}_{2^{m}/2}\left(ax+b\left(\Tr_{2^m/2^k}\left(x^{2q+2}\right)+cx\right)\right)}=\sum_{x\in \mathbb{F}_{2^m}}(-1)^{\mathrm{Tr}_{2^m/2}(Q(x))},
\end{eqnarray*}
where $Q(x):=ax+b\left(\Tr_{2^m/2^k}\left(x^{2q+2}\right)+cx\right)$. 
We now simplify \(\operatorname{Tr}_{2^m/2}(Q(x))\):
\begin{eqnarray*}
&&\mathrm{Tr}_{2^m/2}(Q(x))\\
&=&\mathrm{Tr}_{2^m/2}\left(ax+b\left(\Tr_{2^m/2^k}\left(x^{2q+2}\right)+cx\right)\right)\\
&=&\mathrm{Tr}_{2^m/2}\left(ax+bcx+bx^{2q+2}+\left(b^{\frac{1}{q}}x^{2q+2}\right)^{q}+\cdots+\left(b^{\frac{1}{q^{l-1}}}x^{2q+2}\right)^{q^{l-1}}\right)\\
&=&\mathrm{Tr}_{2^m/2}\left((a+bc)x+\left(b+b^{\frac{1}{q}}+\cdots+b^{\frac{1}{q^{l-1}}}\right)x^{2q+2}\right)\\
&=&\mathrm{Tr}_{2^m/2}\left((a+bc)x+\Tr_{2^m/2^k}\left(b^{\frac{1}{q^{l-1}}}\right)x^{2q+2}\right)\\
&=&\mathrm{Tr}_{2^m/2}\left((a+bc)x+\Tr_{2^m/2^k}\left(b\right)x^{2q+2}\right).
\end{eqnarray*}
If $\mathrm{Tr}_{2^m/2^k}(b)=0$, then we have
\begin{eqnarray*}
W_{f}(a,b)=\sum_{x\in \mathbb{F}_{2^m}}(-1)^{\mathrm{Tr}_{2^m/2}\left((a+bc)x\right)}
=
\begin{cases}
2^m & \text{ if }a+bc=0,\\
0 & \text{ if }a+bc\neq0.
\end{cases}
\end{eqnarray*}
If $\mathrm{Tr}_{2^m/2^k}(b)\neq0$, the bilinear form of $\varphi_{a,b}(x):=\mathrm{Tr}_{2^m/2}(Q(x))$ is given by
\begin{eqnarray*}
&&B_{\varphi_{a,b}}(x,y)\\
&=&\varphi_{a,b}(x+y)+\varphi_{a,b}(x)+\varphi_{a,b}(y)\\
&=&\mathrm{Tr}_{2^m/2}\left(\mathrm{Tr}_{2^m/2^k}(b)x^{2q}y^2+\mathrm{Tr}_{2^m/2^k}(b)x^2y^{2q}\right)\\
&=&\mathrm{Tr}_{2^m/2}\left(\left(\left(\mathrm{Tr}_{2^m/2^k}(b)\right)^{q}y^{2q^2}+\mathrm{Tr}_{2^m/2^k}(b)y^2\right)x^{2q}\right)\\
&=&\mathrm{Tr}_{2^m/2}\left(L_{b}(y)x^{2q}\right),
\end{eqnarray*}
with $L_{b}(y)=\mathrm{Tr}_{2^m/2^k}(b)y^2+\mathrm{Tr}_{2^m/2^k}(b)y^{2q^2}$. 
Denote  by
\begin{eqnarray*}
V_{\varphi_{b}}=\ker(L_b)=\left\{y\in \mathbb{F}_{2^m} : L_{b}(y)=0\right\}.
\end{eqnarray*}
By Lemma \ref{dim}, we have $\dim_{\mathbb{F}_{2}}(V_{\varphi_{b}})=k$.  
Applying Lemma \ref{walsh}, we then deduce that $W_{f}(a,b)\in \left\{0,\pm2^{\frac{m+k}{2}}\right\}$ whenever $\mathrm{Tr}_{2^m/2^k}(b)\neq0$.  
Combining these two cases, we obtain $W_{f}(a,b)\in\left\{0,2^m,\pm2^{\frac{m+k}{2}}\right\}$ for all $a,b\in \mathbb{F}_{2^m}$.

Recall that \(\dim(\mathcal{C}_{f})=2m-\dim_{\mathbb{F}_{2}}(K_{1})\),
where \(K_{1}\) is defined in Equation~\eqref{eq6}.
Note that \(W_{f}(a,b)=2^m\) if and only if \(a+bc=0\) and
\(\operatorname{Tr}_{2^m/2^k}(b)=0\), which yields \(2^{m-k}\) solution pairs
\((a,b)\). Then \(\dim_{\mathbb{F}_{2}}(K_{1})=m-k\) and hence
\(\dim(\mathcal{C}_{f})=m+k\).

By Equation (\ref{eqn-Walsh}), we have 
\begin{eqnarray*}
\operatorname{wt}(\mathbf{c}_{a,b})\in\left\{2^{m-1},\ 0,\ 2^{m-1}-2^{\frac{m+k}{2}-1},\ 2^{m-1}+2^{\frac{m+k}{2}-1}\right\}.
\end{eqnarray*}
According to Lemma \ref{dualdeg3} and the first three Pless power moments, we can easily obtain
the weight distribution of $\mathcal{C}_{f}$ in Table \ref{table1}.

By Lemma \ref{min}, substituting the minimal nonzero weight
$w_{\min}=2^{kl-1}-2^{\frac{kl+k}{2}-1}$ and the maximal nonzero weight
$w_{\max}=2^{kl-1}+2^{\frac{kl+k}{2}-1}$ into Equation \eqref{eq5},
we deduce that $\mathcal{C}_{f}$ is minimal if $k(l-1)\geq4$.
According to Lemma \ref{self}, we further deduce that
$\mathcal{C}_{f}$ is self-orthogonal whenever $k(l+1)\geq 6$.
\end{proof}

\subsection{The case that $f(x)=cx+\Tr_{q^2/q}\left(x^6+x^{2q+4}\right)$}
In this subsection, we study the parameters, weight distribution, minimality, and self-orthogonality of \(\mathcal{C}_{f}\), which is constructed from the function \(f(x)=\operatorname{Tr}_{q^2/q}\bigl(x^6+x^{2q+4}\bigr)+cx\), where \(c\in \mathbb{F}_{q}^{*}\) and \(q=2^{k}\) with \(k\) an odd positive integer. By Lemma \ref{2to1}, this mapping is \(2\)-to-\(1\) over \(\mathbb{F}_{q^2}\) under certain conditions. To present the main results, we require the following lemma.

\begin{lemma}\label{dim2}
Let $m=2k$ and $q=2^k$, where $k$ is an odd positive integer.
 Then we have
\begin{eqnarray*}
\dim_{\mathbb{F}_{2}}(\ker(L_{b}))=k+1,
\end{eqnarray*}
where $L_{b}(y):=\left(b^{2^{k+1}}+b^2\right)y^{2^{k+3}}+\left(b^{2^k}+b\right)y^{2^{k+1}}+\left(b^{2^{k+1}}+b^2\right)y^{2^3}+\left(b^{2^k}+b\right)y^{2}$ is a $2$-linearized polynomial with $b^{2^{k+1}}+b^2\neq 0$ for $y\in \mathbb{F}_{2^m}$.
\end{lemma}

\begin{proof}
 The conventional \(2\)-associate of $L_{b}(y)$ is given by
$l_{b}(y)=\left(b^{2^{k+1}}+b^2\right)y^{k+3}+\left(b^{2^k}+b\right)y^{k+1}+\left(b^{2^{k+1}}+b^2\right)y^{3}+\left(b^{2^k}+b\right)y$. 
According to Lemma \ref{wubaof}, we need to determine
\begin{eqnarray*}
\text{gcrd}\bigl(l_{b}(y),y^{2k}-1\bigr).
\end{eqnarray*}
 Note that $y^{2k}-1=\bigl(y^{k}+1\bigr)^2$. 
Let $A:=b^{2^k}+b$ and $k=2t+1$. Then $l_{b}(y)=A^2y^{k+3}+Ay^{k+1}+A^2y^3+Ay$. 
We have
\begin{align*}
A^2y^{k+3}+Ay^{k+1}+A^2y^3+Ay=A^2y^3\left(y^k+1\right)+Ay^{k+1}+Ay.
\end{align*}
Since the degree of \(Ay^{k+1}+Ay\) is not less than \(k\), we must continue applying the right-hand division algorithm until the degree of the remainder does not exceed \(k-1\). Since
\begin{align*}
Ay^{k+1}+Ay=Ay\left(y^k+1\right),
\end{align*}
we obtain the expression
\begin{align*}
l_b(y)=\bigl(A^2y^3+Ay\bigr)\left(y^k+1\right).
\end{align*}
Hence, we have
$$\text{gcrd}\bigl(l_{b}(y),y^{2k}-1\bigr)=\text{gcrd}\left(\bigl(A^2y^3+Ay\bigr),y^{k}+1\right)(y^k+1).$$
Then it is sufficient to determine $\text{gcrd}\left(A^2y^3+Ay,y^k+1\right)$. 
We have a right-hand division:
\begin{eqnarray*}
y^{k}+1=\bigl(A^{-2}\bigr)^{2^{k-3}}y^{k-3}\bigl(A^2y^3+Ay\bigr)+\bigl(A^{-1}\bigr)^{2^{k-3}}y^{k-2}+1.
\end{eqnarray*}
Similarly, by performing  the right-hand division until the degree of the remainder is less than $3$, we have
\begin{eqnarray}\label{eqn-first}
\nonumber&&y^k+1=\left(\sum_{i=1}^{t}\left(\prod_{j=1}^{i-1}\left(A^{-1}\right)^{2^{k-2j-1}}\right)\left(A^{-2}\right)^{2^{k-2i-1}}y^{k-2i-1}\right)\left(A^2y^3+Ay\right)\\
&&+\left(A^{-1}\right)^{2^{k-3}+\cdots+2^{k-2t-1}}y+1.
\end{eqnarray}
These steps are prone to confusion, so we elaborate the proof in great detail. Denote $2^{k-3}+\cdots+2^{k-2t-1}=2^{\frac{2^{k-1}-1}{3}}=s$, then
\begin{eqnarray}\label{eqn-second}
A^2y^3+Ay&=&\bigl(A^{2+4s}y^2+A^{2+4s}A^{2s}y\bigr)\left(\left(A^{-1}\right)^{s}y+1\right)+\left(A^{2^k}+A\right)y.
\end{eqnarray}
Since $A=b^{2^k}+b$ and $A^{2^k}=b+b^{2^k}$, then we have $A^{2^k}+A=0$. Combining Equations (\ref{eqn-first}) and (\ref{eqn-second}) yields
\begin{eqnarray*}
\text{gcrd}\bigl(A^2y^3+Ay,y^k+1\bigr)=\left(A^{-1}\right)^{s}y+1.
\end{eqnarray*}
Finally, we conclude that 
\begin{eqnarray*}
\text{gcrd}\bigl(l_{b}(y),y^{2k}-1\bigr)=\left(\left(A^{-1}\right)^{s}y+1\right)\bigl(y^k+1\bigr).
\end{eqnarray*}
By Lemma \ref{wubaof}, the desired conclusion follows.
\end{proof}

\begin{theorem}\label{2to1code}
Let $f(x)=\Tr_{q^2/q}\left(x^6+x^{2q+4}\right)+cx\in \mathbb{F}_{q^{2}}[x]$, where $q=2^k$, $k$ is odd, and $c\in\mathbb{F}_{q}^{*}$. Let $m=2k$. Then $\mathcal{C}_{f}$ is a binary linear code with parameters
\begin{eqnarray*}
\bigl[2^{2k}-1,3k,2^{2k-1}-2^{\frac{3k+1}{2}-1}\bigr]
\end{eqnarray*}
and weight distribution in Table \ref{table2}. If $k\geq5$, then $\mathcal{C}_{f}$ is minimal. If $k\geq3$, then $\mathcal{C}_{f}$ is self-orthogonal.
\end{theorem}

\begin{table}[h]
\centering{\caption{Weight distribution of $\mathcal{C}_{f}$ in Theorem \ref{2to1code}}\label{table2}}%
\begin{tabular}{@{}cc@{}}
\hline
Weight & Frequency  \\
\hline
$0$ & $1$ \\
  $2^{2k-1}-2^{\frac{3k+1}{2}-1}$ & $2^{2k-2}+2^{\frac{3k-1}{2}-1}-2^{k-2}-2^{\frac{k-1}{2}-1}$ \\
  $2^{2k-1}$ & $2^{3k}-2^{2k-1}+2^{k-1}-1$ \\
  $2^{2k-1}+2^{\frac{3k+1}{2}-1}$ & $2^{2k-2}-2^{\frac{3k-1}{2}-1}-2^{k-2}+2^{\frac{k-1}{2}-1}$ \\
\hline
\end{tabular}
\end{table}

\begin{proof}
 By Equation (\ref{eqn-Walsh}), we first need to compute the Walsh transforms $W_{f}(a,b)$ for any $a,b\in \mathbb{F}_{2^m}$. We have
\begin{eqnarray*}
W_{f}(a,b)&=&\sum_{x\in \mathbb{F}_{2^m}}(-1)^{\mathrm{Tr}_{2^m/2}\left(ax+b\left(\Tr_{q^2/q}\left(x^6+x^{2q+4}\right)+cx\right)\right)}\\
&=&\sum_{x\in \mathbb{F}_{2^m}}(-1)^{\mathrm{Tr}_{2^m/2}(Q(x))},
\end{eqnarray*}
where $Q(x):=ax+b\left(\Tr_{q^2/q}\left(x^6+x^{2q+4}\right)+cx\right)$. We now simplify $\mathrm{Tr}_{2^m/2}(Q(x))$:
\begin{eqnarray*}
&&\mathrm{Tr}_{2^m/2}(Q(x))\\
&=&\mathrm{Tr}_{2^m/2}\left(ax+b\left(\Tr_{2^m/2^k}\left(x^6+x^{2q+4}\right)+cx\right)\right)\\
&=&\mathrm{Tr}_{2^m/2}\left(ax+bcx+bx^6+bx^{6q}+bx^{2q+4}+bx^{4q+2}\right)\\
&=&\mathrm{Tr}_{2^m/2}\left((a+bc)x+bx^{2^2+2}+b^{2^{k}}x^{2^2+2}+bx^{2^{k+1}+2^2}+b^{2^k}x^{2^{k+1}+2^2}\right)\\
&=&\mathrm{Tr}_{2^m/2}\left((a+bc)x+\left(b^{2^k}+b\right)x^{2^2+2}+\left(b^{2^k}+b\right)x^{2^{k+1}+2^2}\right).
\end{eqnarray*}
If $b^{2^k}+b=0$, then we have
\begin{eqnarray*}
W_{f}(a,b)=\sum_{x\in \mathbb{F}_{2^m}}(-1)^{\mathrm{Tr}_{2^m/2}\left((a+bc)x\right)}=
\begin{cases}
2^m &\text{if }a+bc=0,\\
0& \text{if }a+bc\neq0.
\end{cases}
\end{eqnarray*}
If $b^{2^k}+b\neq0$, the bilinear form of $\varphi_{a,b}(x):=\mathrm{Tr}_{2^m/2}(Q(x))$ is given by
\begin{eqnarray*}
&&B_{\varphi_{a,b}}(x,y)\\
&=&\varphi_{a,b}(x+y)+\varphi_{a,b}(x)+\varphi_{a,b}(y)\\
&=&\mathrm{Tr}_{2^m/2}\left(\left(b^{2^k}+b\right)x^{2^2}y^2+\left(b^{2^k}+b\right)x^2y^{2^2}+\left(b^{2^k}+b\right)x^{2^{k+1}}y^{4}+\left(b^{2^k}+b\right)x^4y^{2^{k+1}}\right)\\
&=&\mathrm{Tr}_{2^m/2}\left(\left(\left(b^{2^k}+b\right)y^{2^{k+1}}+\left(b^{2^{k+1}}+b^2\right)y^{2^{k+3}}+\left(b^{2^{k+1}}+b^2\right)y^{2^3}+\left(b^{2^k}+b\right)y^{2}\right)x^{2^{k+2}}\right)\\
&=&\mathrm{Tr}_{2^m/2}\left(L_{b}(y)x^{2^{k+2}}\right),
\end{eqnarray*}
with $L_{b}(y)=\left(b^{2^k}+b\right)y^{2^{k+1}}+\left(b^{2^{k+1}}+b^2\right)y^{2^{k+3}}+\left(b^{2^{k+1}}+b^2\right)y^{2^3}+\left(b^{2^k}+b\right)y^{2}$. Denote by 
\begin{eqnarray*}
V_{\varphi_{b}}=\ker\bigl(L_{b}\bigr)=\left\{y\in \mathbb{F}_{2^m} : L_{b}(y)=0\right\},
\end{eqnarray*}
By Lemma \ref{dim2}, we have $\dim_{\mathbb{F}_{2}}(V_{\varphi_{b}})=k+1$.  
Applying Lemma \ref{walsh}, we then deduce that $W_{f}(a,b)\in \left\{0,\pm2^{\frac{m+k+1}{2}}\right\}$ whenever $b^{2^k}+b\neq0$.  
Combining these two cases, we obtain $W_{f}(a,b)\in\left\{0,2^m,\pm2^{\frac{m+k+1}{2}}\right\}$ for all $a,b\in \mathbb{F}_{2^m}$.

Recall that \(\dim(\mathcal{C}_{f})=2m-\dim_{\mathbb{F}_{2}}(K_{1})\),
where \(K_{1}\) is defined in Equation~\eqref{eq6}.
Note that \(W_{f}(a,b)=2^m\) if and only if \(a+bc=0\) and
\(\operatorname{Tr}_{2^m/2^k}(b)=0\), which yields \(2^{m-k}\) solution pairs
\((a,b)\). Then \(\dim_{\mathbb{F}_{2}}(K_{1})=m-k\) and hence
\(\dim(\mathcal{C}_{f})=m+k\).

By Equation (\ref{eqn-Walsh}), we have 
\begin{eqnarray*}
\operatorname{wt}(\mathbf{c}_{a,b})\in \left\{2^{m-1},\ 0,\ 2^{m-1}-2^{\frac{m+k+1}{2}-1},\ 2^{m-1}+2^{\frac{m+k+1}{2}-1}\right\}.
\end{eqnarray*}
According to Lemma \ref{dualdeg3} and the first three Pless power moments, we can easily obtain
the weight distribution of $\mathcal{C}_{f}$ in Table \ref{table2}.

By Lemma \ref{min}, substituting the minimal nonzero weight
$w_{\min}=2^{2k-1}-2^{\frac{3k+1}{2}-1}$ and the maximal nonzero weight
$w_{\max}=2^{2k-1}+2^{\frac{3k+1}{2}-1}$ into Equation \eqref{eq5},
we deduce that $\mathcal{C}_{f}$ is minimal if $k\geq5$.
According to Lemma \ref{self}, we further deduce that
$\mathcal{C}_{f}$ is self-orthogonal whenever $k\geq2$.

This completes the proof.
\end{proof}

\subsection{The case that $f(x)=\Tr_{q^l/q}\left(c_{1}x+c_{2}x^{\frac{q+q^2}{2}}\right)+c_{3}x^{2^i}$}
In the subsection, we study the linear code $\mathcal{C}_{f}$ for $f(x)=\Tr_{q^l/q}\left(c_{1}x+c_{2}x^{\frac{q+q^2}{2}}\right)+c_{3}x^{2^i}$, where $i\in \mathbb{Z}$, $c_{1}\in \mathbb{F}_{q^l}$, $c_{2}\in \mathbb{F}_{q}$ and $c_{3}\in \mathbb{F}_{q}^{*}$. In this section, we first prove that $f(x)$ is either a $q$-to-$1$ mapping, a $2^{\gcd(i,k)}$-to-$1$ mapping or a permutation polynomial, and then use $f(x)$ to construct a family of linear codes with three nonzero weights. 

\begin{theorem}\label{q2q2}
Let $q=2^k$ with $k$ an positive integer. Define $f(x)=\Tr_{q^l/q}\left(c_{1}x+c_{2}x^{\frac{q+q^2}{2}}\right)+c_{3}x^{2^i}$, where $i\in \mathbb{Z}$ and $c_{1}\in \mathbb{F}_{q^l}$, $c_{2}\in \mathbb{F}_{q}$, $c_{3}\in \mathbb{F}_{q}^{*}$. The the following statements  hold. 
\begin{enumerate}[(\arabic*)]
  \item If \(\operatorname{Tr}_{q^l/q}(c_1)=lc_2\), then \(f(x)\) is a permutation polynomial over $\mathbb{F}_{q^l}$.
  \item If \(\operatorname{Tr}_{q^l/q}(c_1)\neq lc_2\), then \(f(x)\) is as follows:
  \[
  \begin{cases}
  \text{a \(q\)-to-\(1\) mapping over $\mathbb{F}_{q^l}$} 
  \qquad\qquad\qquad\text{if } i=0 \text{ and } c_3+\operatorname{Tr}_{q^l/q}(c_1)+lc_2=0, & \\[6pt]
  \text{a permutation polynomial over $\mathbb{F}_{q^l}$} 
  \qquad\text{ if } i=0 \text{ and } c_3+\operatorname{Tr}_{q^l/q}(c_1)+lc_2\neq0, & \\[6pt]
  \text{a \(2^{\gcd(i,k)}\)-to-\(1\) mapping over $\mathbb{F}_{q^l}$}  
  \quad\qquad\text{ if } i>0 \text{ and } 
  \left(\dfrac{lc_2+\operatorname{Tr}_{q^l/q}(c_1)}{(c_3)^{1/2^i}}\right)^{\frac{2^m-1}{\gcd(2^i-1,\,2^m-1)}}=1, & \\[10pt]
  \text{a permutation polynomial over $\mathbb{F}_{q^l}$} 
  \qquad\text{ if } i>0 \text{ and } 
  \left(\dfrac{lc_2+\operatorname{Tr}_{q^l/q}(c_1)}{(c_3)^{1/2^i}}\right)^{\frac{2^m-1}{\gcd(2^i-1,\,2^m-1)}}\neq1. &
  \end{cases}
  \]
\end{enumerate}
\end{theorem}

\begin{proof}
We analyze the solutions of the equation $f(x)=e$ for all $e\in \mathbb{F}_{q^l}$.
Let $u:=c_{3}x^{2^i}+e=\Tr_{q^l/q}\left(c_{1}x+c_{2}x^{\frac{q^2+q}{2}}\right)\in \mathbb{F}_{q}$. Then $x=\left(\frac{u+e}{c_{3}}\right)^{\frac{1}{2^i}}$ due to $c_{3}\neq0$. Due to $u,c_{2}\in \mathbb{F}_{q}$, $c_{3}\in \mathbb{F}_{q}^{*}$, we have
\begin{eqnarray*}
x^{\frac{q^2+q}{2}}&=&\left(\frac{u+e}{c_{3}}\right)^{\frac{q^2+q}{2^{i+1}}}=\left(\frac{u^2+\left(e^{q^2}+e^q\right)u+e^{q^2+q}}{\left(c_{3}\right)^2}\right)^{\frac{1}{2^{i+1}}}.
\end{eqnarray*}
Substitute the above expression into $f(x)=e$, we obtain
\begin{eqnarray}\label{eq8}
\nonumber u=&\Tr_{q^l/q}\left(c_{2}\left(\frac{u^2+\left(e^{q^2}+d^q\right)u+e^{q^2+q}}{\left(c_{3}\right)^2}\right)^{\frac{1}{2^{i+1}}}+c_{1}\left(\frac{u+e}{c_{3}}\right)^{\frac{1}{2^i}}\right)\notag\\
\nonumber=&\frac{\Tr_{q^l/q}\left(c_{1}\right)}{\left(c_{3}\right)^{\frac{1}{2^i}}}u^{\frac{1}{2^i}}+\frac{\Tr_{q^l/q}\left(c_{1}e^{\frac{1}{2^i}}\right)}{\left(c_{3}\right)^{\frac{1}{2^i}}}+\frac{lc_{2}}{\left(c_{3}\right)^{\frac{1}{2^i}}}u^{\frac{1}{2^i}}+\frac{c_{2}\left(\Tr_{q^l/q}\left(e^{q^2}+e^q\right)\right)^{\frac{1}{2^{i+1}}}}{\left(c_{3}\right)^{\frac{1}{2^i}}}u+\frac{c_{2}\Tr_{q^l/q}\left(e^{\frac{q^2+q}{2^{i+1}}}\right)}{\left(c_{3}\right)^{\frac{1}{2^i}}}\notag\\
\nonumber =&\frac{lc_{2}+\Tr_{q^l/q}\left(c_{1}\right)}{\left(c_{3}\right)^{\frac{1}{2^i}}}u^{\frac{1}{2^i}}+\frac{\Tr_{q^l/q}\left(c_{1}e^{\frac{1}{2^i}}\right)}{\left(c_{3}\right)^{\frac{1}{2^i}}}+\frac{c_{2}\Tr_{q^l/q}\left(e^{\frac{q^2+q}{2^{i+1}}}\right)}{\left(c_{3}\right)^{\frac{1}{2^i}}}.
\end{eqnarray}
Since $\gcd\left(2^i,2^k-1\right)=1$, the mapping $u\rightarrow u^{2^i}$ is bijective over $\mathbb{F}_{2^k}$ and this transformation preserves the number of roots of the original function. Then rearranging Equation (\ref{eq8}) yields 
\begin{eqnarray}\label{eq9}
\nonumber u^{2^i}&=\frac{lc_{2}+\Tr_{q^l/q}\left(c_{1}\right)}{\left(c_{3}\right)^{\frac{1}{2^i}}}u+\frac{\Tr_{q^l/q}\left(c_{1}e^{\frac{1}{2^i}}\right)}{\left(c_{3}\right)^{\frac{1}{2^i}}}
+\frac{c_{2}\Tr_{q^l/q}\left(e^{\frac{q^2+q}{2^{i+1}}}\right)}{\left(c_{3}\right)^{\frac{1}{2^i}}}.\\
\end{eqnarray}
Now we consider the following cases.

\textbf{Case 1}\quad$i=0$.

Equation (\ref{eq9}) can be rearranged as 
\begin{eqnarray}\label{eq10}
\left(c_{3}+\Tr_{q^l/q}\left(c_{1}\right)+lc_2\right)u+\Tr_{q^l/q}\left(c_{1}e\right)+c_{2}\Tr_{q^l/q}\left(e^{\frac{q^2+q}{2}}\right)
=0.
\end{eqnarray}
Form the above relation, if $c_{3}+\Tr_{q^l/q}\left(c_{1}\right)+lc_2=0$, then for all $u\in\mathbb{F}_{q}$, Equation (\ref{eq10}) is satisfied for $\Tr_{q^l/q}\left(c_{1}e\right)+c_{2}\Tr_{q^l/q}\left(e^{\frac{q^2+q}{2}}\right)=0$, and $f(x)$ admits $q$ distinct solutions. For $\Tr_{q^l/q}\left(c_{1}e\right)+c_{2}\Tr_{q^l/q}\left(e^{\frac{q^2+q}{2}}\right)\neq0$, Equation (\ref{eq10}) never holds and $f(x)$ has no solutions. Then $f(x)$ is a $q$-to-$1$ mapping. If $c_{3}+\Tr_{q^l/q}\left(c_{1}\right)+lc_2\neq0$, Equation (\ref{eq10}) has exactly one solution, hence $f(x)$ is a permutation polynomial of $\mathbb{F}_{q^l}$.

\textbf{Case 2}\quad$i>0$.

While $l$ is an even integer, Equation (\ref{eq9}) can be rearranged as
\begin{eqnarray}\label{eq11}
\left(c_{3}\right)^{\frac{1}{2^i}}u^{2^i}+\Tr_{q^l/q}\left(c_{1}\right)u+\Tr_{q^l/q}\left(c_{1}e^{\frac{1}{2^i}}\right)+c_{2}\Tr_{q^l/q}\left(e^{\frac{q^2+q}{2^{i+1}}}\right)=0.
\end{eqnarray}
According to Lemma \ref{root}, Equation (\ref{eq11}) admits
only two possibilities for its number of solutions in $\mathbb{F}_{2^k}$: either it
has no solution, or its number of solutions equals that of
the homogeneous equation
\begin{eqnarray}\label{eq12}
\left(c_{3}\right)^{\frac{1}{2^i}}u^{2^i}+\Tr_{q^l/q}\left(c_{1}\right)u=0.
\end{eqnarray}
Equation (\ref{eq12}) yields $u=0$ or 
\begin{eqnarray}\label{solution}
u^{2^i-1}=\frac{\Tr_{q^l/q}\left(c_{1}\right)}{\left(c_{3}\right)^{\frac{1}{2^i}}}.
\end{eqnarray}
If $\Tr_{q^l/q}\left(c_{1}\right)=0$, then Equation (\ref{eq12}) has exactly one solution $u=0$. Thus $f(x)$ is a permutation polynomial.
If $\Tr_{q^l/q}\left(c_{1}\right)\neq0$ and $\left(\frac{\Tr_{q^l/q}\left(c_{1}\right)}{\left(c_{3}\right)^{\frac{1}{2^i}}}\right)^{\frac{2^m-1}{\gcd\bigl(2^i-1,2^m-1\bigr)}}=1$, then Equation (\ref{solution}) has $\gcd\left(2^i-1,2^k-1\right)$ distinct nonzero solutions, implying that $f(x)$ is a $2^{\gcd(i.k)}$-to-$1$ mapping. If $\bigl(\frac{\Tr_{q^l/q}\left(c_{1}\right)}{\left(c_{3}\right)^{\frac{1}{2^i}}}\bigr)^{\frac{2^m-1}{\gcd\bigl(2^i-1,2^m-1\bigr)}}\neq1$ and $\Tr_{q^l/q}\left(c_{1}\right)\neq0$, then Equation (\ref{solution}) has no solutions, implying $f(x)$ is a permutation polynomial.

While $l$ is an odd positive integer, Equation (\ref{eq9}) can be rearranged as
\begin{eqnarray}\label{eq13} \left(c_{3}\right)^{\frac{1}{2^i}}u^{2^i}+\left(c_{2}+\Tr_{q^l/q}\left(c_{1}\right)\right)u+\Tr_{q^l/q}\left(c_{1}e^{\frac{1}{2^i}}\right)+c_{2}\Tr_{q^l/q}\left(e^{\frac{q^2+q}{2^{i+1}}}\right)=0.
\end{eqnarray}
According to Lemma \ref{root}, Eqaution (\ref{eq13}) admits
only two possibilities for its number of solutions in $\mathbb{F}_{2^k}$: either it
has no solution, or its number of solutions equals that of
the homogeneous equation
\begin{eqnarray}\label{eq14}
\left(c_{3}\right)^{\frac{1}{2^i}}u^{2^i}+\left(c_{2}+\Tr_{q^l/q}\left(c_{1}\right)\right)u=0.
\end{eqnarray}
Equation (\ref{eq14}) yields $u=0$ or 
\begin{eqnarray}\label{solution2}
u^{2^i-1}=\frac{c_{2}+\Tr_{q^l/q}\left(c_{1}\right)}{\left(c_{3}\right)^{\frac{1}{2^i}}}.
\end{eqnarray}
If $c_{2}+\Tr_{q^l/q}\left(c_{1}\right)=0$, then Equation (\ref{eq14}) has exactly one solution $u=0$. Thus $f(x)$ is a permutation polynomial.
If $c_{2}+\Tr_{q^l/q}\left(c_{1}\right)\neq0$ and $\bigl(\frac{c_{2}+\Tr_{q^l/q}\left(c_{1}\right)}{\left(c_{3}\right)^{\frac{1}{2^i}}}\bigr)^{\frac{2^m-1}{\gcd\bigl(2^i-1,2^m-1\bigr)}}=1$, then Equation (\ref{solution2}) has $\gcd\left(2^i-1,2^k-1\right)$ distinct nonzero solutions, implying $f(x)$ is a $2^{\gcd(i.k)}$-to-$1$ mapping. If $c_{2}+\Tr_{q^l/q}\left(c_{1}\right)\neq0$ and $\left(\frac{c_{2}+\Tr_{q^l/q}\left(c_{1}\right)}{\left(c_{3}\right)^{\frac{1}{2^i}}}\right)^{\frac{2^m-1}{\gcd\bigl(2^i-1,2^m-1\bigr)}}\neq1$, then Equation (\ref{solution2}) has no solutions, implying $f(x)$ is a permutation polynomial.

\end{proof}

\begin{remark}
Theorem \ref{q2q2} assumes that \(l\) is a positive integer, \(c_{1}\in \mathbb{F}_{q^l}\), \(c_{2}\in \mathbb{F}_{q}\), and \(c_{3}\in \mathbb{F}_{q}^{*}\). If \(l\) is odd, \(i=0\), \(c_{2}=1\), \(c_{1}=0\), and \(c_{3}\in \mathbb{F}_{q}\setminus\{0,1\}\), our function \(f(x)\) degenerates into 
\[
f(x)=c_3x+\operatorname{Tr}_{q^{l}/q}\bigl(x^{\frac{q+q^2}{2}}\bigr).
\]
It has been proved in \cite[Theorem 3.5]{qu} that this degenerate \(f(x)\) is a permutation polynomial over \(\mathbb{F}_{q^l}\). However, Theorem \ref{q2q2} does not impose the restriction that \(l\) is odd. Moreover, in addition to permutation polynomials, we also investigate the \(n\)-to-\(1\) mapping property, specifically determining the conditions under which \(f(x)\) is a \(q\)-to-\(1\) mapping or a \(2^{\gcd(i,k)}\)-to-\(1\) mapping.
\end{remark}

Now we study the parameters, weight distribution, minimality, and self-orthogonality of \(\mathcal{C}_{f}\), which is constructed from the \(n\)-to-1 mapping \(f(x)\) in Theorem \ref{q2q2}. To this end, we need the following lemma. 

\begin{lemma}\label{dim3}
Let $m=lk$ and $q=2^k$, where $k,l$ are positive integers. Let $L_{b}(y):=\left(\Tr_{2^m/2^k}(b)c_{2}\right)^2y^{q^3}+\left(\Tr_{2^m/2^k}(b)c_{2}\right)^2y^q$ be a $2$-linearized polynomial with $\Tr_{2^m/2^k}(b)c_{2}\neq 0$ for $y\in \mathbb{F}_{2^m}$. Then we have
\begin{eqnarray*}
\dim_{\mathbb{F}_{2}}(\ker( L_{b}))=
\begin{cases}
k & \mbox{if  $l$ is odd,}\\
2k &  \mbox{if $l$ is even.}\\
\end{cases}
\end{eqnarray*}
\end{lemma}

\begin{proof}
 The conventional \(2\)-associate of $L_{b}(y)$ is given by
$l_{b}(y)=\left(\Tr_{2^m/2^k}(b)c_{2}\right)^2\left(y^{k}+y^{3k}\right)$. 
According to Lemma \ref{wubaof}, we need to determine
\begin{eqnarray*}
\text{gcrd}\left(\left(\Tr_{2^m/2^k}(b)c_{2}\right)^2\left(y^{k}+y^{3k}\right),y^{kl}-1\right).
\end{eqnarray*}
Let $D:=\left(\Tr_{2^m/2^k}(b)c_{2}\right)^2$. Note that $y^{lk}-1=y^{lk}+1$.Then we have a right-hand division:
\begin{align*}
y^{kl}+1=\bigl(D^{-1}\bigr)^{2^{k-3}}y^{k(l-3)}\bigl(Dy^{3k}+Dy^k\bigr)+y^{k(l-2)}+1.
\end{align*}
Since the degree of \(y^{k(l-2)}+1\) is not less than \(3k\), we need to continue applying the right-hand division algorithm until the degree of the remainder does not exceed \(3k-1\).
 When $l$ is odd, we write $l=2t+1$. Similarly to the computational methods in Lemmas \ref{dim} and  \ref{dim2}, we have
\begin{eqnarray*}
y^{kl}+1=\sum_{i=1}^{t}\left(D^{-1}\right)^{2^{k(l-2i-1)}}y^{k(l-2i-1)}\left(Dy^{3k}+Dy^{k}\right)+y^{k}+1
\end{eqnarray*}
and 
\begin{eqnarray*}
Dy^{3k}+Dy^k&=&\left(Dy^{2k}+Dy^k\right)\left(y^k+1\right).
\end{eqnarray*}
Then we conclude that  \[\text{gcrd}\left(\left(\Tr_{2^m/2^k}(b)c_{2}\right)^2\left(y^{k}+y^{3k}\right),y^{kl}-1\right)=y^k+1.\] 

When $l$ is even,  write $l=2t$. We have
\begin{eqnarray*}
y^{kl}+1=\sum_{i=1}^{t-1}\left(D^{-1}\right)^{2^{k(l-2i-1)}}y^{k(l-2i-1)}\left(Dy^{3k}+Dy^{k}\right)+y^{2k}+1
\end{eqnarray*}
and 
\begin{eqnarray*}
Dy^{3k}+Dy^k&=&Dy^{k}\left(y^{2k}+1\right).
\end{eqnarray*}
Then we conclude that \[\text{gcrd}\left(\left(\Tr_{2^m/2^k}(b)c_{2}\right)^2\left(y^{k}+y^{3k}\right),y^{kl}-1\right)=y^{2k}+1.\] 
Consequently, the desired conclusion follows from Lemma~\ref{wubaof}.
\end{proof}

\begin{theorem}\label{q2q2code}
Let $f(x)=\Tr_{q^l/q}\left(c_{1}x+c_{2}x^{\frac{q+q^2}{2}}\right)+c_{3}x^{2^i}\in\mathbb{F}_{q^l}[x]$, where $q=2^k$, $i\in\mathbb{Z}$ and $c_{1}\in \mathbb{F}_{q^l}$, $c_{2}\in \mathbb{F}_{q},c_{3}\in \mathbb{F}_{q}^{*}$. Let $m=kl$. Then the following statements hold.
\begin{enumerate}
\item If $l$ is odd, then $\mathcal{C}_{f}$ is a binary linear code with parameters
\begin{eqnarray*}
\left[2^{lk}-1,lk+k,2^{lk-1}-2^{\frac{lk+k}{2}-1}\right]
\end{eqnarray*}
and weight distribution in Table \ref{table3}. If $k(l-1)\geq4$, then $\mathcal{C}_{f}$ is minimal. If $k(l+1)\geq6$, then $\mathcal{C}_{f}$ is self-orthogonal.
\item If $l$ is even, then $\mathcal{C}_{f}$ is a binary linear code with parameters
\begin{eqnarray*}
\left[2^{lk}-1,lk+k,2^{lk-1}-2^{\frac{lk+2k}{2}-1}\right]
\end{eqnarray*}
and weight distribution in Table \ref{table4}. If $k(l-2)\geq4$, then $\mathcal{C}_{f}$ is minimal. If $k(l+2)\geq6$, then $\mathcal{C}_{f}$ is self-orthogonal.
\end{enumerate}
\end{theorem}

\begin{table}[h]
\centering{\caption{Weight distribution of $\mathcal{C}_{f}$ in Theorem \ref{q2q2code}  (odd $l$)}\label{table3}}%
\begin{tabular}{@{}cc@{}}
\hline
Weight & Frequency  \\
\hline
$0$ & $1$ \\
  $2^{lk-1}-2^{\frac{lk+k}{2}-1}$ & $2^{lk-2}+2^{\frac{lk+k}{2}-2}-2^{lk-k-1}-2^{\frac{lk-k}{2}-1}$ \\
  $2^{lk-1}$ & $2^{lk+k-1}-2^{lk-1}+2^{lk-k}-1$ \\
  $2^{lk-1}+2^{\frac{lk+k}{2}-1}$ & $2^{lk-2}-2^{\frac{lk+k}{2}-2}-2^{lk-k-1}+2^{\frac{lk-k}{2}-1}$ \\
\hline
\end{tabular}
\end{table}

\begin{table}[htbp]
\centering
\caption{Weight distribution of \(\mathcal{C}_{f}\) in Theorem \ref{q2q2code} (even \(l\))}
\label{table4}
\begin{tabular}{cc} 
\hline
Weight & Frequency  \\
\hline
$0$ & $1$ \\
  $2^{lk-1}-2^{\frac{lk+2k}{2}-1}$ & $2^{lk-2}-2^{\frac{lk+2k}{2}-2}-2^{lk-k-1}+2^{\frac{lk-k}{2}-1}$ \\
  $2^{lk-1}$ & $2^{lk+k-1}-2^{lk-k-1}+2^{lk-2k}-1$ \\
  $2^{lk-1}+2^{\frac{lk+2k}{2}-1}$ & $2^{lk-2}+2^{\frac{lk+2k}{2}-2}-2^{lk-k-1}-2^{\frac{lk-k}{2}-1}$ \\
\hline
\end{tabular}
\end{table}

\begin{proof}
By Equation (\ref{eqn-Walsh}), we first need to compute the Walsh transforms $W_{f}(a,b)$ for any $a,b\in \mathbb{F}_{2^m}$. We have
\begin{eqnarray*}
W_{f}(a,b)&=&\sum_{x\in \mathbb{F}_{2^m}}(-1)^{\Tr_{2^m/2}\left(ax+b\left(\Tr_{q^l/q}\left(c_{1}x+c_{2}x^{\frac{q+q^2}{2}}\right)+c_{3}x^{2^i}\right)\right)}\\
&=&\sum_{x\in \mathbb{F}_{2^m}}(-1)^{\Tr_{2^m/2}(Q(x))}.
\end{eqnarray*}
where $Q(x):=ax+b\left(\Tr_{q^l/q}\left(c_{1}x+c_{2}x^{\frac{q+q^2}{2}}\right)+c_{3}x^{2^i}\right)$. 
We now simplify \(\operatorname{Tr}_{2^m/2}(Q(x))\) using the properties of trace functions as follows:
\begin{eqnarray*}
&&\Tr_{2^m/2}\left(Q(x)\right)\\
&=&\Tr_{2^m/2}\left(ax+b\left(\Tr_{q^l/q}\left(c_{1}x+c_{2}x^{\frac{q+q^2}{2}}\right)+c_{3}x^{2^i}\right)\right)\\
&=&\Tr_{2^m/2}\left(ax+\left(bc_{3}\right)^{\frac{1}{2^i}}x+\left(bc_{1}x+b^{\frac{1}{q}}c_{1}x+\cdots+b^{\frac{1}{q^{l-1}}}c_{1}x\right)\right.\\
&&\quad\left.+\left(bc_{2}x^{\frac{q+q^2}{2}}+b^{\frac{1}{q}}c_{1}x^{\frac{q+q^2}{2}}+\cdots+b^{\frac{1}{q^{l-1}}}c_{1}x^{\frac{q+q^2}{2}}\right)\right)\\
&=&\Tr_{2^m/2}\left(\left(a+\left(bc_{3}\right)^{\frac{1}{2^i}}+\Tr_{q^l/q}(b)c_{1}\right)x+\left(\Tr_{q^l/q}(b)c_{2}\right)^2x^{q^2+q}\right).
\end{eqnarray*}
If $\Tr_{q^l/q}(b)c_{2}=0$, then we have
\begin{eqnarray*}
W_{f}(a,b)&=&\sum_{x\in \mathbb{F}_{2^m}}(-1)^{\mathrm{Tr}_{2^m/2}\left(\left(a+\Tr_{2^m/2^k}(b)c_{1}+\left(c_{3}b\right)^{\frac{1}{2^i}}\right)x\right)}\\
&=&
\begin{cases}
2^m & \text{if }a+\Tr_{2^m/2^k}(b)c_{1}+\left(c_{3}b\right)^{\frac{1}{2^i}}=0,\\
0& \text{if }a+\Tr_{2^m/2^k}(b)c_{1}+\left(c_{3}b\right)^{\frac{1}{2^i}}\neq0.
\end{cases}
\end{eqnarray*}
If $\Tr_{q^l/q}(b)c_{2}\neq0$, then the bilinear form of $\varphi_{a,b}(x):=\mathrm{Tr}_{2^m/2}(Q(x))$ is given by
\begin{eqnarray*}
&&B_{\varphi_{a,b}}(x,y)\\
&=&\varphi_{a,b}(x+y)+\varphi_{a,b}(x)+\varphi_{a,b}(y)\\
&=&\Tr_{2^m/2}\left(\left(\Tr_{2^m/2^k}(b)c_{2}\right)^2x^{q^2}y^{q}+\left(\Tr_{2^m/2^k}(b)c_{2}\right)^2x^{q}y^{q^2}\right)\\
&=&\Tr_{2^m/2}\left(\left(\Tr_{2^m/2^k}(b)c_{2}\right)^2\left(y^{q}+y^{q^3}\right)x^{q^2}\right)\\
&=&\Tr_{2^m/2}\left(L_{b}(y)x^{q^2}\right),
\end{eqnarray*}
with $L_{b}(y)=\left(\Tr_{2^m/2^k}(b)c_{2}\right)^2\left(y^{q}+y^{q^3}\right)$. 
Denote  by
\begin{eqnarray*}
V_{\varphi_{b}}=\ker(L_b)=\left\{y\in \mathbb{F}_{2^m} : L_{b}(y)=0\right\}.
\end{eqnarray*}
In what follows, we treat the odd and even cases of \(l\) separately.
\begin{enumerate}
\item By Lemma \ref{dim3}, while $l$ is odd, we have $\dim_{\mathbb{F}_{2}}(V_{\varphi_{b}})=k$.  
Applying Lemma \ref{walsh}, we then deduce that $W_{f}(a,b)\in \left\{0,\pm2^{\frac{m+k}{2}}\right\}$ whenever $\mathrm{Tr}_{q^l/q}(b)c_2\neq0$.  
Combining these two cases, we obtain $W_{f}(a,b)\in\left\{0,2^m,\pm2^{\frac{m+k}{2}}\right\}$ for all $a,b\in \mathbb{F}_{2^m}$.
Recall that \(\dim(\mathcal{C}_{f})=2m-\dim_{\mathbb{F}_{2}}(K_{1})\),
where \(K_{1}\) is defined in Equation~\eqref{eq6}.
Note that \(W_{f}(a,b)=2^m\) if and only if \(a+\Tr_{2^m/2^k}(b)c_{1}+\left(c_{3}b\right)^{\frac{1}{2^i}}=0\) and
\(\operatorname{Tr}_{2^m/2^k}(b)c_{2}=0\), which yields \(2^{m-k}\) solution pairs
\((a,b)\). Then \(\dim_{\mathbb{F}_{2}}(K_{1})=m-k\) and hence
\(\dim(\mathcal{C}_{f})=m+k\).
By Equation (\ref{eqn-Walsh}), we have 
\begin{eqnarray*}
\operatorname{wt}(\mathbf{c}_{a,b})\in\left\{2^{m-1},\ 0,\ 2^{m-1}-2^{\frac{m+k}{2}-1},\ 2^{m-1}+2^{\frac{m+k}{2}-1}\right\}.
\end{eqnarray*}
According to Lemma \ref{dualdeg3} and the first three Pless power moments, we can easily obtain
the weight distribution of $\mathcal{C}_{f}$ in Table \ref{table3}.
By Lemma \ref{min}, substituting the minimal nonzero weight
$w_{\min}=2^{kl-1}-2^{\frac{kl+k}{2}-1}$ and the maximal nonzero weight
$w_{\max}=2^{kl-1}+2^{\frac{kl+k}{2}-1}$ into Equation \eqref{eq5},
we deduce that $\mathcal{C}_{f}$ is minimal if $k(l-1)\geq4$.
According to Lemma \ref{self}, we further deduce that
$\mathcal{C}_{f}$ is self-orthogonal whenever $k(l+1)\geq 6$.

\item By Lemma \ref{dim3}, while $l$ is even, we have $\dim_{\mathbb{F}_{2}}(V_{\varphi_{b}})=2k$.  
Applying Lemma \ref{walsh}, we then deduce that $W_{f}(a,b)\in \left\{0,\pm2^{\frac{m+2k}{2}}\right\}$ whenever $\mathrm{Tr}_{q^l/q}(b)c_2\neq0$.  
Combining these two cases, we obtain $W_{f}(a,b)\in\left\{0,2^m,\pm2^{\frac{m+2k}{2}}\right\}$ for all $a,b\in \mathbb{F}_{2^m}$.
Recall that \(\dim(\mathcal{C}_{f})=2m-\dim_{\mathbb{F}_{2}}(K_{1})\),
where \(K_{1}\) is defined in Equation~\eqref{eq6}.
Note that \(W_{f}(a,b)=2^m\) if and only if \(a+\Tr_{2^m/2^k}(b)c_{1}+\left(c_{3}b\right)^{\frac{1}{2^i}}=0\) and
\(\operatorname{Tr}_{2^m/2^k}(b)c_{2}=0\), which yields \(2^{m-k}\) solution pairs
\((a,b)\). Then \(\dim_{\mathbb{F}_{2}}(K_{1})=m-k\) and hence
\(\dim(\mathcal{C}_{f})=m+k\).
By Equation (\ref{eqn-Walsh}), we have 
\begin{eqnarray*}
\operatorname{wt}(\mathbf{c}_{a,b})\in \left\{2^{m-1},\ 0,\ 2^{m-1}-2^{\frac{m+2k}{2}-1},\ 2^{m-1}+2^{\frac{m+2k}{2}-1}\right\}.
\end{eqnarray*}
According to Lemma \ref{dualdeg3} and the first three Pless power moments, we can easily obtain
the weight distribution of $\mathcal{C}_{f}$ in Table \ref{table4}.
By Lemma \ref{min}, substituting the minimal nonzero weight
$w_{\min}=2^{kl-1}-2^{\frac{kl+2k}{2}-1}$ and the maximal nonzero weight
$w_{\max}=2^{kl-1}+2^{\frac{kl+2k}{2}-1}$ into Equation \eqref{eq5},
we deduce that $\mathcal{C}_{f}$ is minimal if $k(l-2)\geq4$.
According to Lemma \ref{self}, we further deduce that
$\mathcal{C}_{f}$ is self-orthogonal whenever $k(l+2)\geq 6$.
\end{enumerate}

Then the proof is completed. 
\end{proof}

\section{Concluding remarks}\label{sect5}
In this paper, we constructed three families of three-weight binary linear codes from three classes of \(n\)-to-\(1\) mappings. The main contributions are summarized as follows.
\begin{enumerate}
\item We presented two new families of \(n\)-to-\(1\) mappings, which were given respectively by 
\(f(x)=\operatorname{Tr}_{q^l/q}\left(x^{2q+2}\right)+cx\) (Theorem \ref{4to1}) and 
\(f(x)=\operatorname{Tr}_{q^l/q}\left(c_{1}x+c_{2}x^{\frac{q+q^2}{2}}\right)+c_{3}x^{2^i}\) (Theorem \ref{q2q2}). 
\item Using the two newly constructed families of \(n\)-to-\(1\) mappings and one known family, we constructed three families of three-weight projective binary linear codes. Their parameters and weight distributions were explicitly determined via the algebraic theory of skew polynomial rings. Furthermore, we established the minimality and self-orthogonality of these codes.
\end{enumerate}
These binary codes are both minimal and self-orthogonal, which implies their valuable applications in quantum codes and secret sharing schemes with favorable access structures \cite{C, JC}. In addition, our three-weight codes can be used to form association schemes within the framework of \cite{scheme}. Compared with \cite{Lyq,threecode,KC2}, the three-weight codes obtained here possess new parameters. By the weight distributions of our codes and the Pless power moments, one can readily deduce that the duals of these codes have a minimum distance of three. Consequently, these dual codes can be verified to be almost distance-optimal with respect to the well-known sphere-packing bound.

To compute the weight distribution of the linear codes constructed in this paper, the key is to determine the number of zeros of the corresponding linearized polynomials.
From \cite[Page 118]{lidl}, we know that \(G(x)=0\) has \(q^{\ell}\) roots in \(\mathbb{F}_{q^m}\) if and only if
\[
\deg\!\left(\gcd\!\left(\sum_{i=0}^{m-1}\lambda_{i}x^{i},\,x^m-1\right)\right)=\ell,
\]
where \(G(x)=\sum_{i=0}^{m-1}\lambda_{i}x^{q^{i}}\in\mathbb{F}_{q}[x]\) is a \(q\)-linearized polynomial. This result, however, requires \(G(x)\) to have coefficients in \(\mathbb{F}_{q}\), whereas our method based on Lemma \ref{wubaof} applies to the more general case \(G(x)\in\mathbb{F}_{q^m}[x]\). Therefore, this work offers a new perspective for determining the parameters of linear codes derived from \(n\)-to-1 mappings.

\section*{Acknowledgements}
The authors would like to thank Dr. Baofeng Wu for providing Lemma \ref{wubaof}, which facilitated the development of the research methodology in this paper.


 




\vfill

\end{document}